\documentclass[twocolumn,prl,showpacs,amsmath,amssymb]{revtex4}

\usepackage{graphicx}

\def\be{\begin{equation}}
\def\ee{\end{equation}}
\def\bea{\begin{eqnarray}}
\def\eea{\end{eqnarray}}
\def\bma{\begin{mathletters}}
\def\ema{\end{mathletters}}

\def\0{\overline{0}}

\def\q0{\underline{0}}

\def\tr{\mbox{tr}}
\def\one{\leavevmode\hbox{\small1\normalsize\kern-.33em1}}
\def\compl{\begin{picture}(8,8)\put(0,0){C}\put(3,0.3){\line(0,1){7}}\end{picture}}

\def\bra#1{\langle#1|} \def\ket#1{|#1\rangle}

\def\proj#1{\ket{#1}\!\bra{#1}}

\begin{document}

\title{
Quantum correlations and secret bits }

\author{Antonio Ac\'\i n$^{1}$
and Nicolas Gisin$^2$}

\affiliation{ $^1$ICFO-Institut de Ci\`encies Fot\`oniques, Jordi
Girona 29, Edifici Nexus II, E-08034 Barcelona, Spain\\
$^2$GAP-Optique, University of Geneva, 20, Rue de l'\'Ecole de
M\'edecine, CH-1211 Geneva 4, Switzerland }

\date{\today}

%%%%%%%%%%%% Abstract %%%%%%%%%%%%%%%%%%%%%%%%%%%

\begin{abstract}

It is shown that (i) all entangled states can be mapped by
single-copy measurements into probability distributions containing
secret correlations, and (ii) if a probability distribution
obtained from a quantum state contains secret correlations, then
this state has to be entangled. These results prove the existence
of a two-way connection between secret and quantum correlations in
the process of preparation. They also imply that either it is
possible to map any bound entangled state into a distillable probability
distribution or bipartite bound information exists.

\end{abstract}


\maketitle

%\section{Introduction}
%\label{intro}

Entanglement and secret correlations are fundamental resources in
Quantum Information Theory and Cryptography, respectively. They
both share the property of being {\em monogamous} \cite{Terhal},
in the sense that the more two parties share quantum or secret
correlations, the less they are correlated to the outside world.
This fact suggests that these two concepts are closely related.

In the last years, several authors analyzed the link between
quantum and secret correlations. Already in 1991, Ekert
\cite{Ekert} proposed a cryptography protocol whose security was
based on the violation of a Bell inequality \cite{Bell}. More
recently, this link has been exploited for proving the security of
most of the existing quantum cryptography protocols, e.g.
Shor-Preskill proof \cite{SP} of the security of the BB84 scheme
\cite{BB84}. Further relations were later analyzed in
\cite{GW2,CP}. A qualitative equivalence between entanglement and
key distillability has been shown in the cases of two-qubit and of
one-copy distillable states \cite{us}. There even exist
quantitative analogies: the rates of entanglement and of secret
key distillability for some one-way communication protocols are
equal \cite{DW}. All these results suggested the existence of a
correspondence between entanglement and secret key distillability,
in the sense that a quantum state could be transformed into a
private key if and only if it was distillable. However, the recent
results of \cite{HO} have proven this statement to be false: there
are non-distillable quantum states, also known as bound entangled,
that are useful for establishing a secret key.

%Finally, another application of this correspondence
%has been suggested in Ref. \cite{GW2}: the quantum insight might
%be useful for solving open problems in Classical
%Cryptography (see also \cite{CP,CMS}). %the recent and
%surprising results of \cite{HO} have shown that things are
%much subtler than expected and that this correspondence is not as
%straightforward as initially thought.

Up to now, the connection between quantum and secret correlations
has mainly been analyzed from the point of view of distilling or
extracting these resources from quantum states. However, very
little is known about the process of preparation, i.e. about the
resources required for the formation of a quantum state or a
probability distribution. Recall that a state of a composite
quantum system is entangled if and only if it cannot be prepared
by local operations and classical communication (LOCC), that is
{\sl iff} it requires truly quantum correlations (i.e. classical
correlations are not sufficient for its preparation). In a similar
way, for a given probability distribution, one may wonder what the
cost of its distribution is, in terms of secret bits, when only
classical
resources are used. %This corresponds to a scenario where Alice
%and Bob share a public channel and, additionally, have access to
%a common source of secret random bits.
Following \cite{RW}, we say that a probability distribution
contains secret correlations if and only if it cannot be
distributed using only local operations and public communication,
that is {\sl iff} it requires the use of a private channel, (i.e.
public communication is not sufficient for its distribution).

In this work we study those probability distribution derived from
single-copy measurements on bipartite quantum systems. We prove
that (i) all entangled states can be mapped by single-copy
measurements into probability distributions containing secret
correlations and (ii) if a probability distribution containing
secret correlations can be derived from a state $\rho_{AB}$, then
$\rho_{AB}$ has to be entangled. Accordingly, in strong contrast
to the case of distillability, in the preparation process there
exists a one-to-one relation between secret and quantum
correlations. As far as we know, this result represents the first
two-way connection between these two resources, entanglement and
secret correlations. In particular, our results imply that all
bound entangled states are useful to distribute secret
correlations \cite{notedist}, a task that is impossible using only
LOCC. %\cite{notec}.
But let us start reviewing some basic facts
about entanglement and secret correlations.

In the modern theory of quantum correlations, the usual scenario
consists of two parties, Alice and Bob, sharing a quantum state
$\rho_{AB}$ in a system $\compl^{d_A}\otimes \compl^{d_B}$. The
impurity of the state is due to the coupling to the environment.
The basic unit of entanglement is the entangled bit or {\em ebit},
represented by a singlet state, or maximally entangled state of
two qubits, $\ket{\Psi^-}=(\ket{01}-\ket{10})/{\sqrt 2}$. Given
$\rho_{AB}$, one would like to know (i) how many ebits are
required for its preparation and (ii) how many ebits can be
extracted from it by LOCC. These two fundamental questions define
the separability and distillability problems. Associated to them,
there exist two entanglement measures, the so-called Entanglement
Cost, $E_c$, \cite{HHT} and Distillable Entanglement, $E_D$,
\cite{BDSW}. Those states for which $E_c>0$ require ebits for
being prepared, they contain quantum correlations. Separable
states can be prepared by LOCC \cite{Werner}, so $E_c=0$.

Moving to secret correlations, the usual scenario consists of two
honest parties, Alice and Bob, and an eavesdropper, Eve, having
access to independent realizations of three random variables, $X$,
$Y$ and $Z$, characterized by a probability distribution
$P(X,Y,Z)$. Alice and Bob's symbols have some correlations, and
they are also partially correlated with Eve. The basic unit is now
the {\sl secret bit}, that is a probability distribution
$P(X,Y,Z)=P(X,Y)P(Z)$ where $X$ and $Y$ are binary and locally
random, $P(X=Y)=1$ and Eve's symbols are decoupled from Alice and
Bob's result. Similarly as above, given $P(X,Y,Z)$, one can look
for the amount of secret bits (i) needed for its preparation and
(ii) that can be extracted from it by local operations and public
classical communication \cite{GW2,CP}. The corresponding measures
are the so-called Information of Formation, $I_{form}(X;Y|Z)$,
proposed in \cite{RW} as the classical analog of $E_c$, and the
Secret-Key Rate, $S(X;Y||Z)$, introduced in \cite{MW}. As for the
entanglement scenario, a positive information of formation means
that the correlations $P(X,Y,Z)$ cannot be distributed using only
local operations and public communication, secret bits are needed.
Therefore, $P(X,Y,Z)$ contains secret correlations {\sl iff}
$I_{form}(X;Y|Z)>0$ \cite{notedist}.

All these measures, $E_C$ and $E_D$ as well as $S(X;Y||Z)$ and
$I_{form}(X;Y|Z)$, have been defined from an operational point of
view and are hard to compute. For our purpose, it is necessary to
have bounds on these quantities. In the case of secret
correlations, it is known that the so-called intrinsic
information, $I(X;Y\downarrow Z)$, provides a lower bound to the
information of formation \cite{RW} and an upper bound to the
secret-key rate \cite{MW},
\begin{equation}
S(X;Y||Z)\leq I(X;Y\downarrow Z)\leq I_{form}(X;Y|Z) .
\end{equation}
This function, originally introduced in \cite{MW}, is defined as
follows: from $P(X,Y,Z)$ one can compute for any $Z$ the
conditioned probability distribution $P(X,Y|Z)=P(X,Y,Z)/P(Z)$. The
total mutual information between $X$ and $Y$ conditioned on $Z$,
$I(X;Y|Z)$, is the mutual information of $P(X,Y|Z)$ averaged over
all $Z$. The intrinsic information then reads
\begin{equation}
    I(X;Y\downarrow Z)=\min_{Z \rightarrow \bar Z}I(X;Y|\bar Z) ,
\end{equation}
the minimization running over all the channels $Z \rightarrow \bar
Z$.

In order to link quantum and secret correlations, we need two more
remarks.

First, the adversary, Eve, appears in the quantum case in a less
explicit way than in cryptography, where her presence is essential
for the problem to be meaningful. If Alice and Bob share a state
$\rho_{AB}$, the natural way of including Eve is to add a third
system purifying it, in such a way that the global state of the
three parties is
$\ket{\Psi_{ABE}}\in\compl^{d_A}\otimes\compl^{d_B}
\otimes\compl^{d_E}$ and $\rho_{AB}=\tr_E(\proj{\Psi_{ABE}})$.
Thus, all the environment is conservatively associated to the
adversary. Given $\rho_{AB}$, $\ket{\Psi_{ABE}}$ is uniquely
specified up to an irrelevant unitary transformation on Eve's
space \cite{HJW}.

Next, measurements are required for mapping the potential quantum
correlations into probability distributions. We denote by $M_Z$
the positive operators defining Eve's measurement, where $\sum_Z
M_Z=\one_E$, and in a similar way $M_X$ and $M_Y$ define Alice and
Bob's measurements. Thus, given a state $\ket{\Psi_{ABE}}$ and
measurements for each party, the corresponding probability
distribution is
\begin{equation}
\label{prmap}
    P(X,Y,Z)=\tr(M_X\otimes M_Y\otimes M_Z\proj{\Psi_{ABE}}) ,
\end{equation}
while Alice and Bob's probability distribution is
\begin{equation}
\label{prmap2}
    P(X,Y)=\sum_Z P(X,Y,Z)=\tr(M_X\otimes M_Y\rho_{AB}) .
\end{equation}
Notice that the map (\ref{prmap}) is not one-to-one, since there
may be many choices of measurements and states leading to the same
probability distribution. And even if the measurements by Alice,
Bob and Eve are fixed, there may be many states compatible with
Eq. (\ref{prmap}). Therefore, $P(X,Y,Z)$ together with $M_X$,
$M_Y$ and $M_Z$ define equivalence classes in the space of
states $\ket{\Psi_{ABE}}$ \cite{CLL}.% according to Eq.
%(\ref{prmap}).

We have now introduced all the main ideas and can concentrate on
the case where Alice and Bob perform local measurements $M_X$ and
$M_Y$ on an unknown state $\rho_{AB}$. Assume that they can infer
from their data $P(X,Y)$ that the state $\rho_{AB}$ is entangled.
Recall that the detection of entanglement through local
measurements can always be done by means of an entanglement
witness $W$, i.e. by measuring an observable in
$\compl^{d_A}\otimes\compl^{d_B}$ such that for all product states
$\ket{ab}$, $\bra{ab}W\ket{ab}\geq 0$. Recall furthermore that all
operators can be decomposed into a linear combination of product
operators: $W(c_{XY})=\sum_{X,Y} c_{XY}M_X\otimes M_Y$.
Accordingly, whenever a linear combination of Alice and Bob local
measurements provides an entanglement witness $W$, they can
compute its expectation value from their data:
$\tr(W(c_{XY})\rho_{AB})=\sum_{X,Y} c_{XY}P(X,Y)$. And whenever
this expectation value is negative, Alice and Bob can conclude
that the state $\rho_{AB}$ they share is entangled. In such a case
we say that the probability distribution, $P(X,Y)$, for the
measurements $M_X$ and $M_Y$, {\sl is incompatible with any
separable state}. Actually, it was proven in \cite{CLL} that Alice
and Bob can discard any separable state as the origin of the
observed correlations {\sl iff} they can construct from their data
an entanglement witness such that $\tr(W(c_{XY})\rho_{AB})<0$. We
can now state our main result.

Let $\ket{\Psi_{ABE}}$ be a quantum state shared by
Alice, Bob and Eve. The following two statements are equivalent:
\begin{enumerate}
    \item Alice and Bob's state, $\rho_{AB}$, is entangled.
    \item There exist measurements by Alice and Bob, $M_X$
    and $M_Y$, such that for any measurement by Eve, $M_Z$,
    the corresponding probability distribution $P(A,B,E)$
    (\ref{prmap}) contains secret correlations.
\end{enumerate}
This result is indeed a corollary of the following theorem.

{\bf Theorem:} Let $P(X,Y)$ be a probability distribution shared
by Alice and Bob after measuring $M_X$, $M_Y$ on a unknown
state. Then, (i) $P(X,Y)$, for the measurements $M_X$ and $M_Y$, is
incompatible with any separable state (\ref{prmap2}) {\sl if and only
if} (ii) for all the purifications $\ket{\Psi_{ABE}}$, compatible with
the observed data (\ref{prmap2}), and for all measurements
$M_Z$ by Eve, $P(X,Y,Z)$ contains secret correlations.

{\sl Proof:} For the (i) $\Rightarrow$ (ii) part, assume that Alice and
Bob detect the entanglement of the unknown state $\rho_{AB}$ used
for the correlation distribution by means of an entanglement
witness $W(c_{XY})=\sum_{X,Y} c_{XY}M_X\otimes M_Y$ built from
their measurements, i.e. $\tr(W\rho_{AB})<0$. The proof proceeds
by contradiction. Assume that there is a global state
$\ket{\Psi_{ABE}}$ and a measurement by Eve, $M_Z$, such that the
corresponding probability distribution $P(X,Y,Z)=\tr(M_X\otimes
M_Y\otimes M_Z\proj{\Psi_{ABE}})$ admits $P(X,Y)$ as marginal, but
does not contain any secret correlations. This implies
$I(X;Y\downarrow Z)=0$ for $P(X,Y,Z)$, hence there is a channel
$P(\bar Z|Z)$ such that $I(X;Y|\bar Z)=0$, i.e.
\begin{equation}
\label{prfact}
    P(X,Y|\bar Z)=P(X|\bar Z)P(Y|\bar Z) .
\end{equation}
Denote by $\rho_Z$ the state shared by Alice and Bob when Eve's
result is $Z$:
\begin{equation}
    \rho_Z=\frac{1}{P(Z)}\tr_E(\one\otimes M_Z\proj{\Psi_{ABE}}) ,
\end{equation}
where $P(Z)=\tr(\one\otimes M_Z\proj{\Psi_{ABE}})$ and by
$\rho_{\bar Z}$ the state after Eve's classical processing
\begin{equation}
    \rho_{\bar Z}=\frac{1}{P({\bar Z})}\sum_Z P(Z) P(\bar Z|Z)\rho_Z ,
\end{equation}
where $P({\bar Z})=\sum_Z P(Z) P(\bar Z|Z)$ and the positive
operators $M_{\bar Z}=\sum_Z P(\bar Z|Z)M_Z$ define another
measurement, since $\sum_{\bar Z}M_{\bar Z}=\one_E$
\cite{notechannel}. From Eq. (\ref{prfact}) we have that,
$\forall\,X,Y$,
\begin{equation}
\label{stfact}
    \tr(M_X\otimes M_Y\rho_{\bar Z})=\tr(M_X\,\rho_{A\bar Z})
    \tr(M_Y\,\rho_{B\bar Z}) ,
\end{equation}
where $\rho_{A\bar Z}$ ($\rho_{B\bar Z}$) denotes the state after
tracing Bob (Alice) out in $\rho_{\bar Z}$. Define the separable
state $\rho_{AB}^S=\sum P({\bar Z})\rho_{A\bar
Z}\otimes\rho_{B\bar Z}$. Using that $\rho_{AB}=\sum_Z
P(Z)\rho_Z=\sum_{\bar Z} P({\bar Z})\rho_{\bar Z}$, it follows
from Eq. (\ref{stfact}) that
\begin{equation}
    \tr(W\rho_{AB})=\tr(W\rho_{AB}^S)<0 ,
\end{equation}
which is a contradiction with the assumption that $W$ is an
entanglement witness. Therefore, $I_{form}(X;Y|Z)\geq
I(X;Y\downarrow Z)>0$ for all the states $\ket{\Psi_{ABE}}$ and
all Eve's measurements. This conclude the (i) $\Rightarrow$ (ii)
part of the proof.

For the (ii) $\Rightarrow$ (i) part, we proceed again by contradiction
(see also \cite{GW2,CLL}). Assume that there exists a separable
state $\rho_{AB}$ compatible with the observed data $P(X,Y)$.
Since $\rho_{AB}$ is separable, it can be expressed as
\begin{equation}
    \rho_{AB}=\sum_{Z=1}^{n_Z} P(Z)\proj{a_Z\,b_Z} .
\end{equation}
Consider the purification
\begin{equation}
    \ket{\Psi_{ABE}}=\sum_{i=1}^{n_Z} \sqrt{P(Z)}
    \ket{a_Z\,b_Z}\ket{Z} ,
\end{equation}
where $\ket{\Psi_{ABE}}\in\compl^{d_A}\otimes\compl^{d_B}
\otimes\compl^{n_Z}$ and $\ket{Z}$ are $n_Z$ orthonormal vectors.
If Eve applies the measurement defined by $M_Z=\proj{Z}$, we have
that for all Alice and Bob's measurements
\begin{eqnarray}
    P(X,Y|Z)=\tr(M_X\otimes M_Y\proj{a_Z\,b_Z})=\phantom{asdfg}&&\nonumber\\
    \tr(M_X\proj{a_Z})\tr(M_Y\proj{b_Z})=P(X|Z)P(Y|Z).&&
\end{eqnarray}
Now, it is clear that these correlations could be as well
generated using public communication. The random variables $X$ and
$Y$ are locally generated according to one of the $n_Z$
probability distributions $P(X|Z)$ and $P(Y|Z)$. The choice among
these probability distributions is made according to the
probability distribution $P(Z)$. Alice and Bob correlate this
choice through the message $Z$ that one of the parties generates
and sends to the other through a public channel (or a source to
both parties). This classical message is accessible to Eve. No
secret bits are required for this distribution, thus $I_{form}=0$,
which contradicts statement (ii). Hence there is no separable
state $\rho_{AB}$ compatible with the observed data $P(X,Y)$.
$\Box$

{\bf Corollary:} Consider an entangled state
$\rho_{AB}=\tr_E(\proj{\Psi_{ABE}})$, where $\ket{\Psi_{ABE}}$
denotes the global state including Eve. There always exist
measurements by Alice and Bob mapping this state into a
probability distribution containing secret correlations,
independently of Eve's measurement.

{\em Proof:} This easily follows from the previous Theorem
together with two known results: (i) any entangled state is
detected by an entanglement witness \cite{LKCH} and (ii) any
entanglement witness can be decomposed in terms of tensor product
of operators defining local measurements, i.e. can be computed by
local measurements \cite{GHBELMS}. $\Box$

These statements prove the announced ``if and only if" connection
between secret and quantum correlations in the process of
preparation. %Notice that there exist measurements which are state
Note that all the proofs have been derived using single-copy
measurements. This fact allows to easily translate our conclusions
from entanglement based to prepare and measure protocols using the
same ideas as in \cite{BBM} (see also \cite{CLL}). In the
following lines, several implications of the results are
discussed.

First, %it has to be stressed that
the presented connection is as
strong as it could be. Consider that Alice and Bob are connected
by an unknown quantum channel (share an unknown state). As soon as
their measurements outcomes detect that the channel allows one to
distribute entanglement, they know to share secrecy, no matter
what Eve does \cite{notedist}. Alice and Bob may not have enough
information from the obtained measurement results for completely
determine their channel. Still, if their data are only compatible
with entanglement, the observed distribution contains secret bits.
%Thus, an alternative preparation of the same correlations using
%only classical means is not possible without consuming this
%resource.
On the other hand, if the measured correlations are
compatible with a separable state, no secret key can be extracted
from them \cite{CLL}. Indeed, from the observed data Alice and Bob
cannot exclude that $I(X;Y\downarrow Z)=0$, which implies
$S(X;Y||Z)=0$. Thus, any entangling channel can be seen as a
source of privacy.

Next, all this discussion is independent of the distillability
properties of quantum states. Indeed, the previous corollary
provides a systematic way of mapping bound entangled states into
probability distribution containing secrecy. An interesting open
question is whether all these probability distributions are
distillable into a perfect key (c.f. \cite{HO}). It follows from
our results that at least one of the two following possibilities
must be true \cite{notebound} (i) all bound entangled states can
be mapped into distillable probability distributions or (ii) there
exist classical probability distributions having non-distillable
secret correlations%, i.e. their secrecy content is bound
. In this case, they would provide examples of bipartite
probability distributions with bound information, the
cryptographic analog of bound entanglement conjectured in
\cite{GW2} (see also \cite{ACM}).

Finally, our results also shed light on %another open question in
%Quantum Information Theory:
what the differences between bound entangled states and the set of
LOCC operations are. %(or separable states)?
It is known that these states do not improve the fidelity of
teleportation compared to LOCC \cite{HHH}. On the other hand,
there are examples of bound entangled states violating some
inequalities for variance of observables that are satisfied by
separable states \cite{var}. Moving to secret correlations, there
exist bound entangled states that are useful for key distribution
\cite{HO}. Here, it is proven that all bound entangled states can
be transformed into probability distributions that, from the point
of view of its secrecy features, can never be established by LOCC.
%\cite{notec}.

To conclude, in this work we have shown the correspondence between
secret and quantum correlations in the process of preparation.
Given a probability distribution, its formation using quantum
resources needs entangled states if and only if an alternative
preparation using classical resources requires secret bits.

\medskip

%\section{Acknowledgements}

This work is supported by the Swiss NCCR, ``Quantum Photonics" and
OFES within the European project RESQ (IST-2001-37559), the
Spanish MCyT, under ``Ram\'on y Cajal" grant, and Generalitat de
Catalunya.


\begin{references}
\bibitem{notedist}
Whether these secret correlations can be extracted or distilled
into perfect secret bits, a key, is a different, though related,
problem that we do not consider here.
\bibitem{notechannel}
Note that the classical post-processing, $P(\bar Z|Z)$, can always
be included in Eve's measurement.
\bibitem{notebound}
Both possibilities can be true since: (i) one can derive many
probability distribution containing secret correlations from any
bound entangled state and (ii) bound information could be proven
for probability distributions not associated to quantum states.
\end{references}
\end{document}